# AI-Augmented Inquiry and Regulation in Hybrid Systems: A Control Allocation Architecture for Preserving Epistemic Agency in Hybrid Human-AI Cognition

Jochen Kuhn[1], Peter Gerjets[2,3], Ulrich Trautwein[3], Jeffrey A. Greene[4], Sarah Malone[5], Patrik Vogt[6], & Tim Fütterer[3]

*[1]Ludwig-Maximilians-Universität München, Munich, Germany*
*[2]Leibniz-Insitut für Wissensmedien, Tübingen, Germany*
*[3]University of Tübingen, Tübingen, Germany*
*[4]The University of North Carolina at Chapel Hill, USA*
*[5]Saarland University, Saarbrücken, Germany*
*[6]Heidelberg University of Education, Heidelberg, Germany*

**Abstract**

Generative artificial intelligence (genAI) systems are increasingly integral to epistemic processes such as hypothesis generation, explanation construction, and decision-making. Although these systems reliably enhance performance, emerging evidence reveals a metacognitive dilemma: as external generative capacity increases, internal monitoring, calibration, and cognitive engagement may decline. This reflects a redistribution of cognitive control within distributed human-AI systems that cannot be explained by automation bias or reliance on algorithms alone. We propose the AIRIS (AI-Augmented Inquiry and Regulation in Hybrid Systems) framework to analyze this dilemma and specify where regulatory intervention can counteract it. AIRIS is a multi-level control allocation architecture specifying the conditions under which epistemic agency can be preserved in hybrid generative systems. Drawing on distributed cognition, cognitive load theory, multimedia learning, and self-regulated learning, it identifies seven interacting mechanisms through which hybrid cognition may become destabilized, from delegation and calibration drift to motivational-affective drift, and five regulatory operators (Anticipate, Interrogate, Reflect, Integrate, and Synthesize) that target internal generative engagement at points of emerging instability. The architecture does not itself improve learning; it specifies what must remain in place for genAI-supported work to sustain understanding, whether through instructional design, teacher guidance, or learners' own regulation. From it we derive testable propositions concerning the seven mechanisms and the regulatory effects of the five operators, reframing AI augmentation as a problem of control allocation in distributed generative systems. Beyond theory, AIRIS provides a research agenda, a design framework for genAI-integrated learning environments, and a conceptual toolkit for navigating the governance of hybrid human-AI cognition.

## 1. Introduction

### 1.1 The Performance-Metacognition Dilemma in Hybrid Cognition

Consider a graduate student writing a scientific review article. She uses a large language model (LLM) to generate a first draft of her argument, refines the phrasing, and submits work that reads fluently and scores well on surface-quality metrics. Yet when later asked to explain or defend her core argument, she struggles: the text was coherent and polished, but the reasoning was not hers. This scenario illustrates the central challenge we address: as people use generative artificial intelligence (genAI) systems to participate in epistemic activities, generating arguments, explanations, and models - output quality may improve whereas humans' engagement and understanding quietly erode. Although our examples come primarily from learning contexts, the argument applies wherever people sustain epistemic reasoning over time, including professional knowledge work, scientific inquiry, and expert judgment. What is at stake is not learning in the narrow sense but the capacity to generate and evaluate knowledge claims oneself.

The intuitive promise of genAI is clear: a professional who offloads routine drafting, information retrieval, and initial structuring can focus on higher-order reasoning, evaluation, and judgment, remaining the epistemic agent while the AI functions as a powerful assistant. This rests on an implicit distinction we make explicit: delegating truly peripheral, routine work (i.e., *outsourcing*) is generally unproblematic, whereas delegating the learning-relevant generative and regulatory processes themselves (i.e., *cognitive offloading* in the sense at issue here) is precisely what threatens durable understanding. Throughout this paper, our concern is with cognitive offloading in the latter, learning-relevant sense, not with the outsourcing of some routine tasks. A recent metacognitive-threshold account similarly distinguishes productive from harmful offloading (Guerrero-Chirinos et al., 2026); AIRIS differs by specifying a broader mechanism-to-operator architecture for dynamic control allocation across the human–AI interaction cycle.

genAI systems are transforming epistemic practices across education, research, medicine, engineering, and knowledge work (Birhane et al., 2023; Bommasani et al., 2021; Kasneci et al., 2023). Empirical studies have documented productivity gains when individuals collaborate with LLMs on argumentative text, structured code, and other externalized reasoning (e.g., Brynjolfsson et al., 2023; Dell'Acqua et al., 2023; Noy & Zhang, 2023), fueling widespread narratives of cognitive augmentation. Yet emerging evidence suggests a more complex picture.

genAI-supported performance gains do not uniformly correspond to improved metacognitive calibration and durable understanding. In some contexts, individuals using genAI perform better on immediate tasks while exhibiting weaker confidence-accuracy alignment and impaired independent generative reasoning, that is, their ability to construct, evaluate, and defend arguments without genAI (Bastani et al., 2025; Ekin, 2026). This dissociation is what we call the performance-metacognition dilemma: genAI-supported work can improve while the constructive processing underlying durable expertise erodes. However, many findings on genAI and learning are complicated by methodological concerns; a recent meta-analysis of genAI effects on cognitive learning outcomes in STEM found the evidence base fragmented and methodologically heterogeneous (Boolzen et al., 2026). Weidlich et al. (2025) argued that much research reporting genAI learning benefits measured the wrong outcome: the performance *while* the genAI tool is still available, and treating this tool-assisted performance as evidence of learning, rather than testing what the learner can do independently, once the tool is withdrawn. [1] Yan et al. (2025) sharpened this distinction further, arguing that *performance gains* and *genuine learning* must be analytically disaggregated: high output quality under genAI support does not warrant inferring that durable, transferable knowledge has been acquired. Large-scale evidence further shows that reflective and hybrid genAI designs can support delayed AI-free transfer more effectively than direct AI feedback, suggesting that durable learning depends on preserving active evaluative work by learners (Ates, 2026). Converging evidence comes from randomized controlled trials by Liu et al. (2026): although genAI assistance improved performance while available, participants subsequently performed worse on their own and were more likely to abandon tasks, indicating that immediate support can come at the cost of independent competence and persistence.

[1] Throughout this paper, we distinguish between assisted task performance (output quality during genAI-supported work) and transfer performance (durable knowledge and competence demonstrated in subsequent tasks without genAI access). The performance-metacognition dilemma refers specifically to the dissociation between these two: gains in assisted performance do not necessarily correspond to gains in transfer performance.

What is at stake is *epistemic agency* as the capacity to initiate, monitor, validate, coordinate, and take responsibility for one's own knowledge-generating processes.. Recent accounts accordingly frame preserving human epistemic agency as a central challenge of learning with genAI (Wu et al., 2025; Yan et al., 2025).

Particularly, the fluency of genAI outputs may lead users to trust their surface quality and to cede control over the epistemic product without participating in its generation or revision. We argue , however, that the key phenomenon is not trust in genAI per se, but the redistribution of cognitive control within distributed generative systems. The concern of this paper is not that genAI seemingly performs generative epistemic work in the human's place, but that its outputs are treated as if they were the product of such work, and that this impression deepens with use. Two consequences follow: uncritical acceptance yields weaker epistemic products than the surface fluency suggests and, more centrally, forecloses the learning that performing the generative work oneself would have produced. What is required is therefore not merely engagement but epistemic vigilance: a sustained readiness to evaluate genAI outputs rather than adopt them, preserving a division of labor in which the human remains the generative and evaluative authority. This is the capacity that the mechanisms outlined below place at risk. When learners treat genAI outputs as finished products rather than as material to be evaluated and reconstructed, internal constructive engagement, monitoring one's own understanding, integrating new information into existing schemas, and investing effort in reasoning, may gradually vanish.

**1.2 Toward a Regulatory Account of Hybrid Cognition**

If the performance–metacognition dilemma is real, a fundamental question arises: what would have to be true for genAI-supported work to sustain rather than erode understanding? This sets up the central theoretical problem that this paper addresses: *how can epistemic agency be preserved when generative capacity is increasingly externalized to artificial agents?* Existing theories illuminate parts of this problem but, as we argue in Section 2, none of them answers it directly. To address this gap, we propose AIRIS (AI-Augmented Inquiry and Regulation in Hybrid Systems), as a framework that (a) describes how cognitive control is redistributed in human-AI systems, (b) identifies normative conditions under which such redistribution preserves rather than erodes epistemic agency, and (c) derives testable propositions and design principles for regulated hybrid cognition. The benchmark against which AIRIS evaluates hybrid cognition is epistemic agency, not efficiency. GenAI reliably raises productivity and immediate output quality, but whether these gains correspond to durable understanding and responsible knowledge production depends on whether learners retain meaningful control over the generative processes that produce knowledge. Hybrid systems must therefore be judged not by what they produce but by whether such control remains learner-held. This normative claim is complemented by an empirical one: the components of epistemic agency connect to measurable constructs that already show genAI-related degradation under unregulated use (e.g., Bastani et al., 2025; Abdelghani et al., 2026). Related work similarly argues for a principled division of epistemic labour that retains reflective control with humans (Woodruff & Hewitt, 2026); AIRIS extends this position by specifying the recurrent mechanisms through which control is destabilized and the regulatory operators through which it can be restored during human–AI interaction. Recent evidence similarly shows that reactive versus proactive agent designs redistribute conversational initiative and may create an "agency gap" when system initiative is misaligned with learner capability (Jin et al., 2026); AIRIS extends this interaction-level account by specifying the mechanisms through which control is destabilized and the operators through which it can be restored. A related line of work reconceptualizes genAI as a dialogic or epistemic partner rather than an answer provider, emphasizing questioning, justification, and the coordination of multiple perspectives (Deguchi et al., 2026; Tang & Putra, 2026). AIRIS is complementary in specifying how cognitive control is allocated within such interactions and how it can be regulated when it drifts. We use the term framework, rather than model, to signal that AIRIS is integrating several theoretical traditions and yields testable propositions and design principles, rather than specifying a single closed set of explanatory relations. Figure 2 provides an overview of the framework as a whole, showing how the seven destabilizing mechanisms (the diagnostic layer) and the five regulatory operators (the regulatory layer that acts on them) jointly describe where epistemic agency is at risk and specify how it can be sustained; readers may find it useful as an advance organizer for the sections that follow. AIRIS reframes genAI use not as assistance alone but as a regulatory problem: preserving epistemic agency requires specifying how cognitive control is allocated and stabilized across internal and external generative processes. This framing aligns with recent calls to design AI in education to build learner agency rather than dependency (Fütterer, Steinhäuser, et al., 2026; Fütterer, Gerjets, et al., 2026).
The article is structured as follows (see Fig. 1).

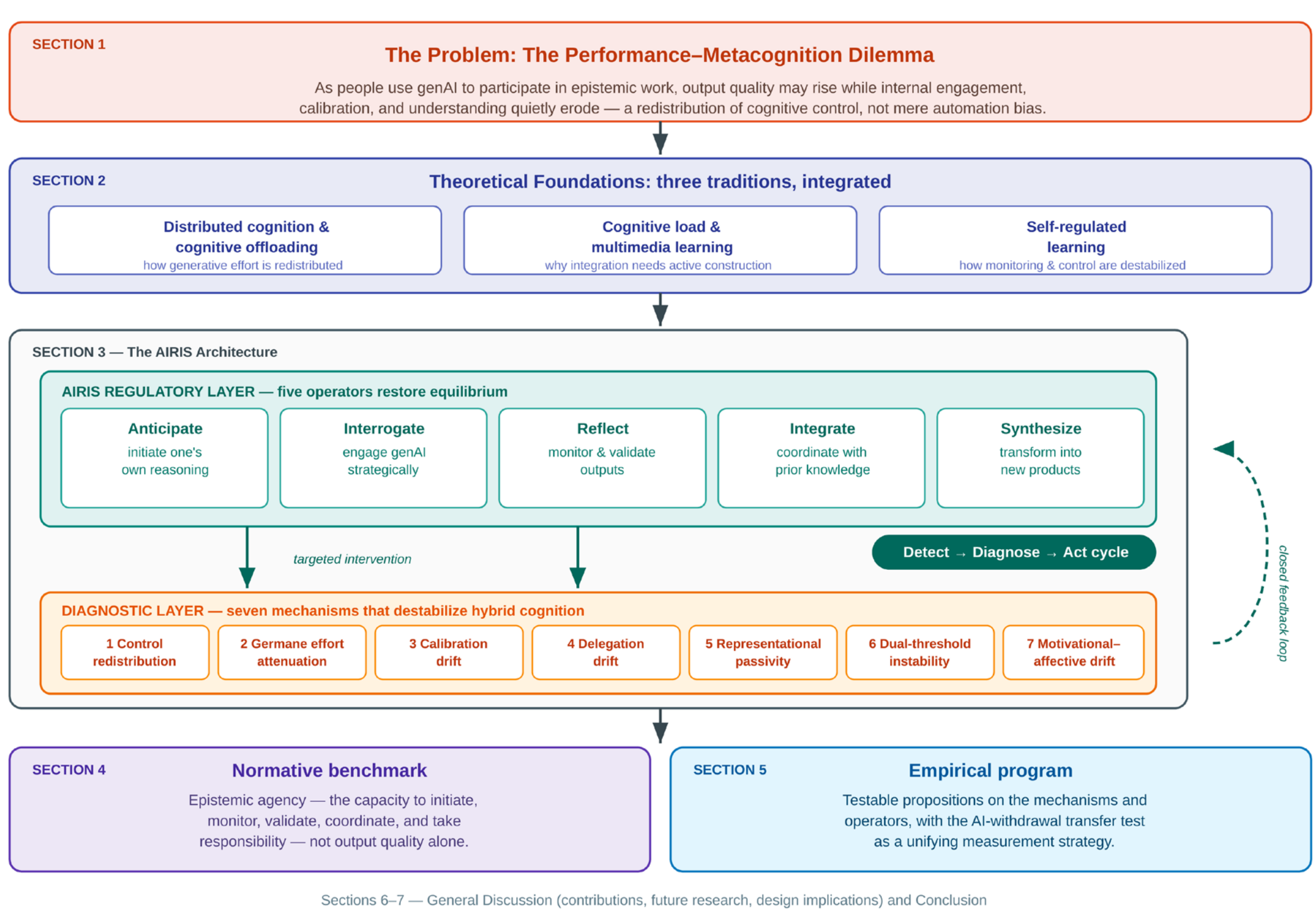

Fig. 1: Graphical advance organizer for the article. The figure traces the argument from the performance–metacognition dilemma (Section 1) through the three integrated theoretical traditions (Section 2) to the AIRIS architecture (Section 3), in which five regulatory operators act on seven destabilizing mechanisms via a Detect–Diagnose–Act cycle, and finally to epistemic agency as the normative benchmark (Section 4) and the empirical program (Section 5).

After setting out the performance–metacognition dilemma that motivates the framework in Section 1, Section 2 develops the theoretical foundations by integrating three previously largely separate traditions: resource-based accounts (cognitive load theory, distributed cognition, and cognitive offloading), representation-based theories (multimedia learning and multiple external representations), and self-regulated learning. Section 3 introduces the AIRIS architecture: seven mechanisms through which hybrid cognition becomes unstable, five regulatory operators (Anticipate, Interrogate, Reflect, Integrate, Synthesize) that restore equilibrium, and a Detect–Diagnose–Act cycle linking behavioral traces to targeted interventions. Section 4 positions epistemic agency as the normative benchmark for evaluating hybrid cognition, arguing that output quality alone is insufficient. Section 5 derives testable propositions that constitute a multi-level empirical research program. Finally, Section 6 discusses theoretical contributions and design implications, and Section 7 concludes.

# 2. Theoretical Foundations

## 2.1 Three Theoretical Traditions for Analyzing Hybrid Cognition

Three research traditions have each addressed part of how generative cognition is distributed. Regulation-based research has developed sophisticated models of self-regulated learning (SRL; Greene, 2017; Greene et al., 2024; Winne & Hadwin, 1998; Zimmerman, 2000). Resource- and representation-based research has modeled the integration of different information sources under resource limitations (cognitive load theory [CLT], the cognitive theory of multimedia learning [CTML], and multiple external representations [MERs]). Research on distributed cognition and cognitive offloading, in turn, examines how generative effort is distributed across human-(Clark & Chalmers, 1998; Dellermann et al., 2019; Järvelä et al., 2023, 2025; Risko & Gilbert, 2016). Yet these traditions rarely converge in research on genAI: each captures one facet of the problem -SRL the regulatory dynamics, CLT/CTML/MERs the representational and resource demands, distributed cognition and cognitive offloading the division of labor between human and machine. However, none integrates resource allocation, representational integration, and regulatory monitoring into a single account of how control is redistributed when an external agent participates in generation. It is this missing integration, rather than gaps in understanding of any single mechanism, that leaves current frameworks unable to specify when genAI-supported work sustains epistemic agency and when it erodes it. Empirical work has begun to clarify mediating mechanisms. For example, Alam et al. (2026) reported that AI-assisted learning tools can improve outcomes when they successfully maintain student engagement while keeping cognitive load within manageable limits, a finding that underscores the critical mediating role of generative engagement. Converging evidence points in the same direction from the failure end: studies of unregulated genAI use document a learning–performance dissociation in which assisted output quality rises whereas independent competence stagnates or declines (Bastani et al., 2025). Field studies of inquiry with conversational genAI have shown learners frequently fail to regulate their questioning and instead accept fluent but shallow answers (Abdelghani et al., 2026). Together, these findings indicate that the same tool can either sustain or erode generative engagement depending on how the interaction is structured - precisely the contingency that a regulatory account must explain. However, the conditions under which genAI promotes versus undermines such engagement remain insufficiently theorized, and the architecture of cognitive control redistribution in hybrid systems has yet to be systematically addressed.

Understanding hybrid human-AI cognition requires integrating theoretical traditions that have historically evolved in parallel. Distributed cognition proposes that cognitive processes are not confined to the individual mind but extend into artifacts, tools, and social systems, a view that treats the entire cognitive system as the unit of analysis , including both internal processes and external resources (Clark & Chalmers, 1998; Hutchins, 1995). A canonical example is the ship's navigation team: the knowledge of where the vessel is and where it is heading resides not in any single crew member but in the distributed system of charts, instruments, roles, and communication (Hutchins, 1995). Within this framework, cognition emerges through interactions between internal representations and external supports. Cognitive offloading research has demonstrated individuals routinely externalize cognitive demands to artifacts and environments to reduce internal processing load (Risko & Gilbert, 2016). Offloading may target memory (e.g., outsourcing retrieval to external storage; Sparrow et al., 2011), but also extends to generative operations such as planning, reasoning, and argument construction. genAI represents an especially powerful target for generative operations: unlike a notebook or search engine, it does not merely store or retrieve information but actively produces new content. Crucially, cognitive offloading is not a passive reflex but the outcome of a metacognitive decision. Learners weigh - though rarely deliberately - whether the cognitive effort of internal processing outweighs the benefits of externalizing a task. The decision is metacognitive in that it draws on judgments about one's own capabilities and effort costs, not in that it is consciously deliberated (Dunn & Risko, 2016). This metacognitive decisional layer is essential for understanding how genAI destabilizes hybrid cognition: genAI's convenience systematically biases this cost-benefit calculation toward externalization. Here the distinction introduced earlier becomes decisive: externalizing routine, learning-incidental operations (i.e., outsourcing) is unproblematic, whereas externalizing the generative operations that constitute the learning itself (i.e., offloading in the learning-relevant sense) is what destabilizes hybrid cognition. Nothing about genAI compels this; rather, the perceived reduction of effort costs makes problematic externalization the path of least resistance. The distinction is empirically consequential, not merely conceptual. In Liu et al. (2026), participants with AI access who used it primarily to obtain direct solutions (61%) showed declining independent performance relative to their own pretest, whereas those who used it for hints and clarification (27%) showed improved performance. The groups did not differ at pretest, indicating that the divergence arose from how the tool was used rather than from who used it. Thus, genAI extends cognitive offloading beyond memory support toward epistemic production. The critical issue is not whether offloading occurs, but how the distribution of generative effort between internal cognition and external systems affects learning and reasoning.

CLT is valuable for understanding this outsourcing vs. offloading distinction precisely because it already implies when delegation should help (i.e., offloading that removes extraneous load is beneficial) and the framework specifies the resource logic for why. What it does not specify is how learners should allocate generative effort once an external agent can perform constructive work as well, and it is this gap that the distributed, generative setting exposes. With that affordance and limit in view, CLT provides an essential analytical lens for understanding why these distribution patterns matter for learning. Working memory is severely limited, and effective learning environments must manage cognitive resource allocation carefully (Sweller, 1988; Sweller et al., 2019). CLT distinguishes among intrinsic load (e.g., task complexity), extraneous load (e.g., inefficient design), and constructive processing (i.e., processing that contributes to schema formation). Whereas earlier formulations treated this as a third load type ("germane load"), Sweller (2023) has argued since that it is better understood as resource allocation rather than a separate source of load; we adopt the latter view and speak of constructive processing throughout (closely related to what Mayer terms generative processing in the multimedia-learning tradition; Mayer, 2021). genAI can reduce working-memory demands by handling operations that a learner has already automated, or that are peripheral to the learning goal: routine retrieval, formatting, structural scaffolding. Note that these are not "extraneous" in the technical sense of poor instructional design; they are legitimate task components. Many generative operations that genAI can perform, however, such as drafting arguments or constructing explanations, are constructive: they constitute the constructive processing that drives learning. When genAI takes over such tasks, it does not reduce extraneous load but displaces constructive engagement. Nonetheless, constructive processing, the effort required for schema construction, remains essential. If genAI outputs replace rather than scaffold this engagement, learning depth may decline even as surface performance improves. The contrast between replacing and scaffolding is a simplification: in practice, genAI interaction is often iterative, and learners set their own stopping criteria, continuing until an output seems good enough, or until they judge that they have understood enough. Where that threshold falls, and what governs it, is itself a regulatory question, and one to which we return in Section 4.

A second theoretical lens sharpens this concern by focusing on how multiple representations are processed and combined. Multimedia learning and representational coordination bring this into view. The CTML holds that meaningful learning requires active integration of verbal and visual representations (Mayer, 2021). The Design, Functions, Tasks (DeFT) framework for learning with multiple external representations similarly emphasizes that representations support learning only when learners actively translate and coordinate them (Ainsworth, 2006). GenAI dramatically increases the quantity and variety of representational availability, yet abundance does not guarantee integration. When a learner asks genAI to translate between formats, an equation into a graph, a paragraph into a diagram, it returns the coordinated result directly, removing the translation friction that, in the DeFT account, drives the construction of an integrated mental model (Ainsworth, 2006). Then, the learner receives the product of coordination without performing it. This displacement is not inevitable, it arises only when representational coordination is delegated rather than attempted, but the ease of delegation makes it a standing risk. In such cases the observable output is a well-integrated set of representations, whereas the internal integration that the output is supposed to reflect may never occur. GenAI representations may create an illusion of understanding (Abdelghani et al., 2026) that masks the absence of deeper conceptual organization in both internal and external representations.

A third tradition, SRL theory, explains how individuals regulate and coordinate these learning processes over time. At its core, SRL theory identifies monitoring (i.e., tracking one's own understanding and progress) and control (i.e., adjusting strategies in response to monitoring signals) as the mechanisms that drive effective learning (Winne & Hadwin, 1998). An additional critical element is goal clarity: the precision with which learners define what they want to achieve (i.e., durable understanding vs. immediate performance) fundamentally shapes which regulatory processes are engaged; recent experimental work shows that genAI support can strengthen these self-regulatory processes when it is designed to scaffold rather than replace them (Fütterer et al., 2026). In collaborative contexts, regulation may extend beyond the individual[2]: genAI introduces a new *participant* (i.e., an agent that actively contributes regulatory and generative acts, proposing strategies, monitoring apparent progress, generating content, rather than merely serving as a passive tool or information store) into this regulatory loop. An AI teammate can suggest strategies, monitor apparent progress, and generate explanations, the very functions that SRL frameworks assign to the learner. This marks a shift from a descriptive claim, what genAI can and cannot reliably do, to a normative one.

Whereas the preceding paragraphs identified descriptive limitations, what genAI cannot reliably do, the following point is normative: ultimate evaluative responsibility must remain with the human, regardless of how capable the system becomes. genAI bears no responsibility for the epistemic commitments implied by these outputs. That means that genAI cannot be held responsible for whether a generated explanation is correct, whether a proposed strategy is appropriate, or whether the user has actually understood the reasoning it produced. This matters beyond accuracy. If learning and correct output were the only concerns, a sufficiently reliable genAI system would be an acceptable substitute for human evaluation. Yet, genAI is not an acceptable substitute because epistemic commitments require a bearer: someone must be answerable for whether a claim is warranted. Therefore, human agents retain evaluative authority not merely because genAI is fallible, but because accountability cannot be delegated to a system that cannot hold it. Consequently, any account of how hybrid cognition should be regulated must preserve human evaluative authority even when AI teammates are substantially involved in generative processes.

[2] For a more extended treatment of collaborative regulation in GenAI-mediated settings, including phenomena such as responsibility diffusion and epistemic co-construction, see Section 4 and Proposition 6.

Together, these three research traditions (i.e., distributed cognition, cognitive load, and self-regulated learning) play complementary rather than interchangeable roles here, and it is worth stating them explicitly. Distributed cognition and cognitive-offloading research characterize how generative labor is redistributed across the human-AI system; cognitive load theory and the CTML/DeFT tradition characterize how representations must be actively integrated for that redistribution not to hollow out understanding; and self-regulated learning theory characterizes the monitoring and control processes by which a learner could, in principle, govern the partnership. Taken together, they provide the conceptual vocabulary for understanding what happens when genAI becomes a participant in epistemic work. In the next section, we identify the specific mechanisms through which this participation can destabilize the human side of the cognitive partnership.

## 2.2 How Hybrid Cognition Becomes Unstable

Hybrid human-AI cognition constitutes a distributed generative system in which epistemic functions are partially externalized to artificial agents. We use the term epistemic equilibrium to denote a state in which the division of generative and regulatory labor between learner and genAI is balanced such that the learner retains sufficient generative engagement, monitoring, and evaluative authority to sustain durable understanding. This contrasts with a drift in which learning-relevant work is progressively offloaded (in the sense distinguished earlier from the mere outsourcing of routine tasks). The qualifier “productive” matters here: a hybrid system can settle into a stable but unproductive state, in which low engagement and high reliance reinforce one another without any felt disruption, so stability alone is not the criterion; the criterion is durable understanding, which depends on sustaining the learner’s own generative participation. The theories and evidence reviewed above jointly suggest that this equilibrium can be disturbed in identifiable ways; the following seven mechanisms specify those disturbances as theoretically derived, empirically testable hypotheses, not as established facts, nor as a claim that disturbance is inevitable. The seven mechanisms point to specific processes in which human-AI collaboration stops being generatively productive and begins to erode the epistemic robustness it was meant to support. Each is grounded in existing theoretical frameworks and supported by convergent empirical evidence, but they should be understood as hypotheses about processes likely to operate across many hybrid cognition contexts, not as verified universal laws. Their derivation draws on the three theoretical traditions introduced above: what distributed cognition theory implies about offloading, what CLT and CTML predict about schema formation and representational integration, and what SRL theory predicts about the conditions for effective monitoring and control. These convergent predictions concern the epistemic equilibrium of hybrid cognition. Equilibrium here is a structural claim, not a metaphor: productive human-AI collaboration depends on maintaining sufficient levels of human generative engagement and AI reliability (i.e., the degree to which genAI outputs are accurate and appropriately calibrated for the task at hand). When either drops below a critical threshold, the system can compensate. When both drop simultaneously, the result is not merely additive failure but a characteristic nonlinear breakdown, a pattern we term *dual-threshold instability*.

Next, we describe the seven mechanisms in turn, beginning with the one from which the others follow: (1) *Control redistribution* is the foundational mechanism. GenAI systems perform epistemic operations that were once internal to human reasoning (e.g., drafting arguments, proposing hypotheses, organizing reasoning chains). From a distributed cognition perspective, this constitutes a redistribution of generative effort across the cognitive system as a whole (Clark & Chalmers, 1998). Within SRL frameworks, this redistribution shifts the concern: the problem is not that genAI regulates the learner poorly, but that it can remove the perceived need for regulation altogether: a learner who types a question and receives a complete answer has no occasion to plan an approach, monitor comprehension, or evaluate progress. The regulatory cycle is not performed badly; it is not entered (Winne & Hadwin, 1998). Crucially, control redistribution is not inherently detrimental: cognitive offloading regarding planning, for instance, could also free resources for monitoring (Risko & Gilbert, 2016). The critical issue is whether some kind of internal generative engagement remains active or becomes progressively displaced. Darvishi et al. (2024) provided direct behavioral evidence for potential problems resulting from outsourcing core SRL processes to genAI: in genAI-assisted peer-feedback tasks, students who had access to genAI feedback showed measurable reductions in independent evaluative reasoning, a redistribution pattern that persisted and deepened across repeated interactions. This pattern suggests that control displacement is not a one-time adaptation but a progressive reconfiguration of epistemic agency.

(2) *Germane effort attenuation* follows from unmanaged redistribution. In CLT terms, constructive processing refers to the cognitive effort invested in schema construction, in actively organizing, elaborating, and integrating new information into existing knowledge structures. This is the load that drives learning. When learners construct explanations themselves, they necessarily engage in schema-building. When instead they evaluate a fluent, pre-constructed genAI explanation, the demand shifts from construction to comprehension and evaluation: the structure is given, and the learner needs only to assess whether it seems correct. This shift reduces constructive processing precisely where it is most needed. Consistent with this account, Stadler et al. (2024) reported that LLM support reduces learners' experienced mental effort while compromising the depth of their scientific inquiry, cognitive ease purchased at the cost of constructive engagement. If learners primarily receive and recognize genAI outputs rather than generating, reconstructing, or reformulating knowledge themselves, internal conceptual models will remain shallow even if the genAI output is sophisticated (Chi, 2009; Fiorella & Mayer, 2016; Mayer, 2021; Renkl, 2011). Germane effort attenuation, however, does not mean genAI necessarily reduces learning; rather, it highlights the importance of maintaining structured generative engagement even when genAI increases efficiency.

(3) *Calibration drift* describes the decoupling of confidence judgments from actual understanding. GenAI outputs are typically linguistically fluent and structurally coherent, creating an impression of comprehension even when internal understanding remains incomplete. Unlike classical automation bias, which concerns misplaced trust in algorithmic outputs (Parasuraman & Riley, 1997), calibration drift reflects reduced monitoring of one's own understanding rather than of the genAI output per se. When a genAI explanation is fluent and coherent, processing it feels effortless, and that ease can be misread as depth of understanding -a genAI-specific instance of the illusion of fluency documented in metacognition research (Bjork et al., 2013). Then, the impression can substitute for genuine comprehension checking. Learners may fail to ask themselves whether they could reconstruct the argument without the genAI, whether they can identify gaps, or whether they could defend the reasoning under scrutiny. Reflection phases are compressed not because learners trust the genAI blindly, but because the genAI's fluency makes self-assessment feel unnecessary. Calibration also shapes how strongly reliance develops. Thus, confidence-accuracy coupling may weaken progressively - not through deliberate neglect but through the gradual erosion of reflective habits. Recent experimental work has documented precisely such metacognitive deterioration in unrestricted GenAI-use conditions (Bastani et al., 2025). In K. Bauer et al. (2026), participants who happened to observe the system err early relied on it substantially less thereafter and showed no post-removal skill deficit relative to controls, whereas those who did not observe early errors performed worse. Exposure to system fallibility functioned as a calibration correction, one that occurred by chance rather than by design. Fan et al. (2024) term this dynamic *metacognitive laziness*: the convenience of genAI outputs undermines the planning, monitoring, and revision processes central to SRL, not through deliberate avoidance but through a gradual erosion of reflective engagement. Field evidence showed middle-school learners failing to regulate inquiry with conversational genAI, accepting fluent answers without adequate evaluation (Abdelghani et al., 2026), a pattern that has prompted explicit calls to design genAI so that it fosters learner agency rather than dependency (Fütterer, Gerjets, et al., 2026; Fütterer, Steinhäuser, et al., 2026). Zhang and Xu (2025) identified a corresponding paradox of self-efficacy: genAI use increases task-related confidence while intensifying technological dependence, such that individuals feel more capable precisely as their independent generative competence erodes.

(4) *Delegation drift* is the temporal counterpart of control redistribution. Whereas Mechanism 1 describes how work is divided within a given episode, delegation drift describes how that division shifts across episodes: the point at which genAI is consulted moves progressively earlier, until consultation precedes rather than follows independent reasoning. Experimental evidence for this temporal pattern comes from outside genAI contexts. K. Bauer et al. (2026) gave participants access to a machine-learning decision aid across repeated rounds and then removed it without warning. Participants followed the system's predictions increasingly often over time without learning when to overrule them: the probability of overriding an incorrect prediction declined across rounds, and post-removal performance was significantly lower than in a control group that never had access. That this pattern appeared with predictive rather than generative systems suggests delegation drift is not specific to genAI, but genAI's fluency and scope plausibly amplify it. Early interactions may involve consulting genAI only after independent reasoning attempts, whereas later interactions begin with genAI consultation. Cognitive offloading research demonstrates that individuals adaptively externalize tasks to minimize effort (Risko & Gilbert, 2016). In generative contexts, such externalization may extend to reasoning processes themselves. Over time, this can affect self-efficacy beliefs: learners who consistently arrive at answers through genAI may come to attribute competence to the tool rather than to themselves.

(5) *Representational passivity* concerns not how much constructive processing occurs, but of what kind: whereas Mechanism 2 addresses the amount of schema-building effort invested, representational passivity addresses the coordination between representational formats: translating a verbal explanation into a diagram, reading an equation as a graph, or reconstructing a model in one's own notation. A learner may process a genAI explanation thoroughly and still never perform this translation. It occurs when users accept externally coherent genAI representations without actively reconstructing them internally. Research on multiple external representations has shown learning benefits arise when individuals translate and coordinate representations rather than merely viewing them (Ainsworth, 2006). If genAI diagrams or explanations are inspected but not reconstructed, internal representational networks may remain weakly integrated. Paradoxically, the abundance of genAI representations may mask rather than resolve conceptual fragmentation. The critical difference from static learning materials (e.g., illustrated textbooks) lies not in the nature of the representation itself but in the production dynamic: genAI generates individualized, on-demand representations that precisely target the user's query, removing the need to translate across multiple sources or construct one's own organizational structure. This bespoke availability reduces the effort required to produce a coherent-looking representation, without requiring the integrative work that produces genuine understanding. For instance, Yang et al. (2025) used learning analytics on 1,445 writing sessions and found that students who passively accepted genAI text into their own work without reconstructing, paraphrasing, or critically revising it produced work of declining quality over time. This constitutes a representational passivity effect in the domain of written knowledge production: what matters is not which representation is used but whether the learner actively processes and transforms it.

(6) *Dual-threshold instability* differs in kind from the mechanisms described so far: whereas Mechanisms 1 through 5 describe processes by which equilibrium erodes, dual-threshold instability specifies the condition under which that erosion process grows nonlinearly. It describes a pattern of disproportionate performance degradation - not a gradual decline but an accelerated breakdown that occurs specifically when both internal generative engagement and genAI reliability simultaneously fall below the demands imposed by task complexity. When individuals remain actively engaged, moderate genAI errors can often be detected and corrected. Conversely, highly reliable genAI outputs may compensate for reduced internal engagement in well-structured tasks. Instability emerges when both sources of epistemic control become insufficient simultaneously, causing errors to increase disproportionately because neither human monitoring nor genAI reliability provides adequate compensation. This mechanism reframes hybrid failure as a breakdown of complementarity rather than a simple accumulation of independent errors, aligning with hybrid intelligence theory (Dellermann et al., 2019). Whereas direct experimental tests of the dual-threshold prediction remain scarce, indirect support comes from Bastani et al. (2025), whose experimental designs revealed that performance degradation under genAI use was most severe precisely in the condition combining low internal engagement with high genAI fluency. This pattern was consistent with the joint-insufficiency prediction rather than a simple main effect of either variable.

(7) *Motivational-affective drift* describes the gradual recalibration of effort norms as genAI reduces the friction associated with complex tasks. Individuals may become accustomed to rapid solution generation and experience reduced motivation to engage in demanding generative reasoning. Even when cognitive load remains manageable, individuals may choose lower levels of engagement when high-quality outputs are obtainable with minimal effort. Such shifts in effort valuation may amplify other destabilizing mechanisms, particularly delegation drift and constructive-processing attenuation, creating self-reinforcing spirals of disengagement. An additional motivational dimension concerns reduced ownership: when outputs are largely produced by a genAI system, individuals may feel diminished authorship and creative agency. Research on intrinsic motivation suggests that the experience of competence (i.e., having mastered a difficult challenge through one's own effort) is a primary source of autonomous motivation (Ryan & Deci, 2000; Bandura, 1997). If high-quality outputs can be obtained with minimal effort, the opportunities for such mastery experiences diminish, potentially eroding the intrinsic motivation that sustains long-term engagement with demanding epistemic work. Motivation in learning, moreover, is not driven solely by mastering challenges but also by the act of learning itself, for instance, the satisfaction of closing one's own knowledge gaps (Bardach & Murayama, 2025). To the extent that genAI forecloses the experience of identifying and resolving such gaps oneself, this additional, curiosity-based source of motivation is placed at risk, also. Direct evidence for this mechanism now exists. In the withdrawal experiments described above (Liu et al., 2026), participants who had worked with AI did not merely perform worse once it was removed. They gave up more often, skipping problems at roughly twice the rate of controls despite explicit assurance that wrong answers carried no penalty. Because skipping was costless, it reflected a deliberate decision not to engage. The authors attributed this to a shifted reference point for how much effort a task should require: each act of offloading makes unaided work feel comparatively more costly, which in turn makes future offloading more attractive, a self-reinforcing dynamic of precisely the kind described here.

Taken together, these seven mechanisms constitute a map of where and how hybrid cognition can lose its equilibrium. That is, they point to specific processes in which the collaboration between humans and genAI ceases to be generatively productive and begins to erode the epistemic robustness it was meant to support. Crucially, these mechanisms are not independent failure modes but form a self-reinforcing system. These seven mechanisms interact: control redistribution initiates structural shifts in generative engagement; constructive processing attenuation reduces constructive depth; calibration drift weakens monitoring coherence; delegation drift alters long-term problem-solving strategies; representational passivity limits conceptual integration; dual-threshold instability explains nonlinear performance breakdown; and motivational drift reduces voluntary engagement (see Mechanism 7: motivational-affective drift). For instance, delegation drift feeds motivational-affective drift, which accelerates attenuation of constructive processing, which in turn deepens calibration drift and increases vulnerability to dual-threshold breakdown. Initial evidence for sequential dependence comes from Kırcaburun (2026), who found across two independent samples that the association between genAI use and critical thinking dispositions was carried entirely by indirect pathways, through metacognitive weakness, through epistemic laziness, and through a sequential path linking the two, with no detectable direct effect. The cross-sectional design precluded causal inference, but the pattern was consistent with cascading rather than independent effects. It is precisely because these dynamics compound and interact that they cannot be addressed by isolated design decisions. They require a regulatory architecture that can detect when and where the system is drifting, diagnose which mechanisms are active, and intervene in a targeted manner to restore productive equilibrium. Thus, the seven mechanisms do not merely describe instability; they define the problem space that any adequate regulatory framework must address. It is worth emphasizing that these mechanisms describe conditions of potential destabilization, not inevitable outcomes. When learners maintain active, generative engagement, consult genAI in a targeted manner, and systematically evaluate genAI outputs, hybrid cognition can be genuinely productive, achieving quality and depth that neither humans nor genAI could achieve alone. For instance, when learners actively engage with rather than passively accept genAI output, writing quality, experienced ownership, and the writing experience itself can improve (N. Bauer et al., 2026; Yang et al., 2025). The equilibrium point, where human agency and genAI support are mutually reinforcing rather than substitutive, is precisely what AIRIS aims to characterize and maintain.

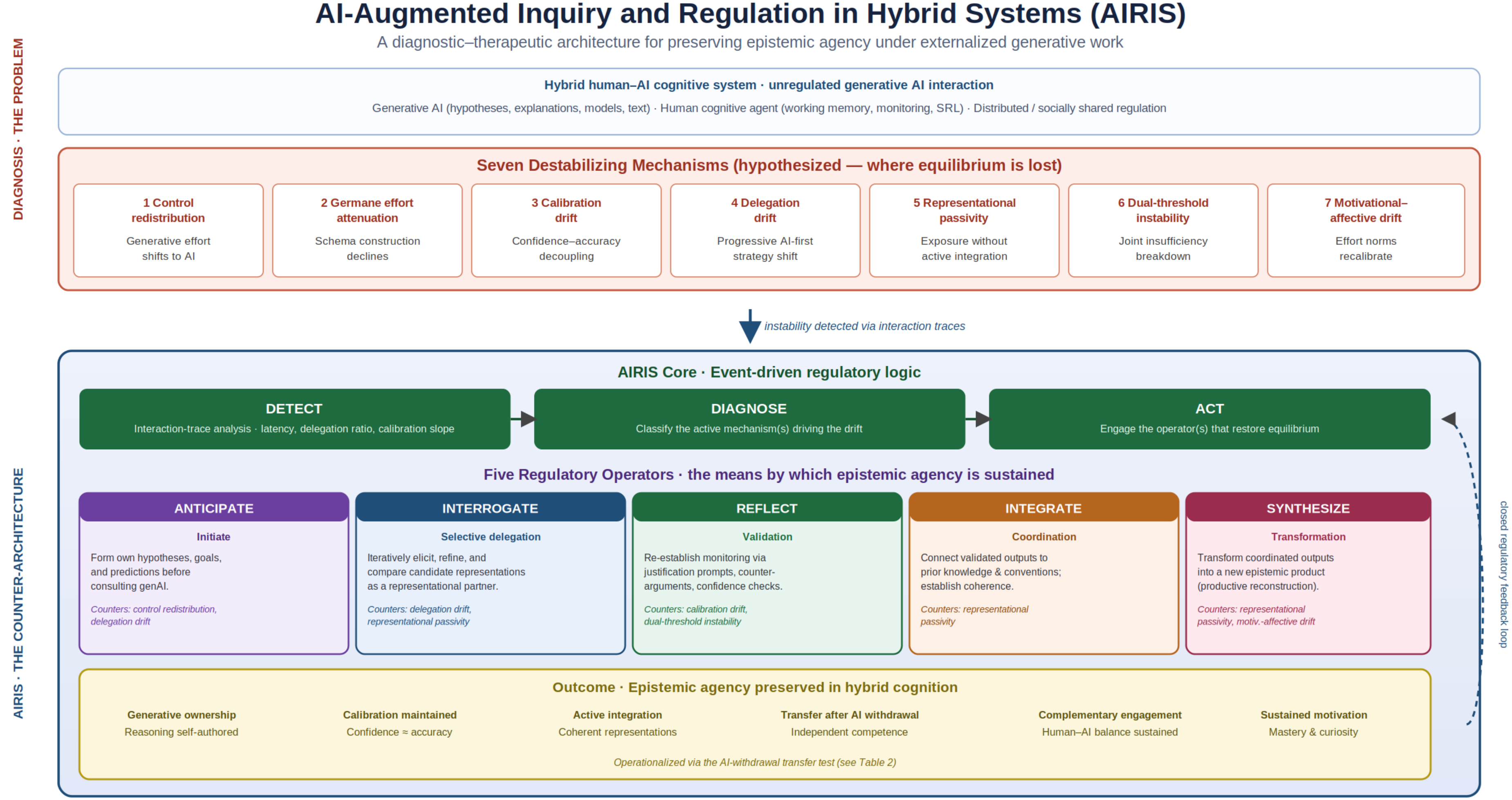

Fig. 2: Schematic overview of the AIRIS framework, illustrating the seven destabilizing mechanisms, the five regulatory operators, and the Detect-Diagnose-Act cycle that links them. The figure is organized as two layers: a diagnostic layer (the seven mechanisms, indicating where hybrid cognition loses equilibrium) and the AIRIS regulatory layer that acts above it (the Detect-Diagnose-Act core and the five operators, Anticipate, Interrogate, Reflect, Integrate, and Synthesize, with their functions of initiation, selective delegation, validation, coordination, and transformation). Each operator is annotated with the mechanism(s) it primarily counteracts (see Table 1), and the preserved-agency outcomes are operationalized through the AI-withdrawal transfer test (see Table 2). The dashed arrow denotes the closed regulatory feedback loop from outcomes back to detection.

## 3. The AI-Augmented Inquiry and Regulation in Hybrid Systems Architecture

### 3.1 Core Principles and Regulatory Logic

AI-Augmented Inquiry and Regulation in Hybrid Systems (AIRIS) is a theoretical framework for understanding how to stabilize distributed generative systems. AIRIS is not a cognitive architecture in the sense of ACT-R or SOAR (Anderson et al., 2004; Laird, 2012): it does not model how the hybrid system functions, but rather specifies what would have to be in place to counteract the drift such systems otherwise exhibit. The term architecture here denotes a structured regulatory framework, a theoretically grounded set of principles, mechanisms, and intervention points, rather than a technical system or a software implementation. The AIRIS acronym was first used in a practice-oriented three-phase classroom framework (Activate–Inquire–Reflect with Intelligent Support; Kuhn et al., 2026a;b); the present account develops that precursor into the five-operator control-allocation architecture specified here. AIRIS is normatively oriented in its benchmark, preserving epistemic agency, but empirically testable in its mechanisms and operators. Rather than prescribing a fixed instructional sequence, it specifies the regulatory capabilities required for genAI-supported work to sustain understanding. AIRIS serves as an episodic control-allocation layer that activates when destabilizing patterns emerge. Its function is not to constrain genAI usage but to restore equilibrium between human generative engagement and genAI-supported processes whenever the balance drifts. Therefore, a primary competence that AIRIS is meant to cultivate is the ability to discriminate between situations in which delegating to genAI is appropriate (i.e., outsourcing of learning-incidental work) and those in which it is not (i.e., offloading of the generative work that constitutes the learning itself).

The aim is not to minimize genAI use but to keep this discrimination, and the evaluative authority it presupposes, with the human agent. AIRIS rests on three core principles. Their justification derives from the theoretical analysis in Section 2: epistemic primacy is warranted because generative engagement is necessary for durable understanding (Mechanisms 1 and 2); complementary generative engagement reflects the insight that human and genAI capabilities are most valuable when they scaffold rather than substitute each other (Mechanism 4). Event-driven intervention is preferable to continuous restriction because instability arises at specific thresholds rather than continuously (Mechanism 6), so that support can preserve the efficiency benefits of genAI while activating regulatory support only when genuine instability signals emerge.

First, human epistemic primacy: Although genAI may generate hypotheses, explanations, or candidate solutions, final evaluative authority remains with the human agent. Second, complementary generative engagement: genAI augments internal generative reasoning rather than replacing it. Third, event-driven intervention: regulatory support is engaged when detectable markers signal emerging instability, rather than continuously or according to predetermined schedules. AIRIS specifies when and why such support should be engaged; whether this is enacted by a teacher, by the learner’s own internalized regulation, or by a future technical implementation is left open. One of AIRIS's most consistent design implications is the principle , an application of the Anticipate operator: wherever possible, humans should engage in at least minimal independent reasoning before consulting genAI. This implication is consistent with the evidence, being directly supported by Pi et al. (2026): when learners identified errors in their own work before receiving genAI feedback, they achieved substantially larger learning gains than when the genAI identified the errors for them, indicating that the sequence of engagement, own reasoning first, genAI second, matters independently of the feedback content. This sequencing is not arbitrary, it ensures that the human agent enters the genAI consultation with a cognitive stake in the outcome, increasing the likelihood that genAI outputs are actively evaluated rather than passively accepted.

Situating AIRIS within existing regulation theory clarifies its logic. In classical SRL frameworks, learners cycle through forethought, performance, and reflection phases (Zimmerman, 2000). GenAI tends to intervene most strongly within the performance phase by supplying explanations, code, and arguments. This assistance increases efficiency but risks compressing the forethought and reflection phases that, even without genAI, are demanding and often only partially realized. AIRIS restores balance by ensuring that forethought processes (e.g., goal articulation, hypothesis formation) and reflective processes (e.g., evaluation, calibration) remain active even when genAI provides substantial performance support.

A similar logic applies to co-regulation. In the learning sciences, co-regulation in the SRL sense refers to external regulatory support provided by a more capable partner, such as a teacher, tutor, or peer, that sustains regulatory processes the learner cannot yet maintain alone (Hadwin et al., 2018). This is distinct from shared regulation, in which peers jointly coordinate regulatory activity (see also Järvelä & Hadwin, 2013). genAI does not fit this description: it can suggest strategies or prompt reflection, but it is not a more capable partner in the required sense, because it cannot take responsibility for the regulatory process or adjust its support to the learner developmental trajectory. It is better understood as a conditional scaffold than as a co-regulating partner (Hadwin et al., 2018). Yet genAI differs crucially from human collaborators: it bears no responsibility for epistemic commitments. AIRIS therefore maintains genAI as a conditional scaffold rather than an epistemic authority, ensuring human agents re-engage in evaluation and integration processes to preserve epistemic accountability.

### 3.2 Event-Driven Triggers and the Five Operators

The operators are episodic in their activation (i.e., triggered by specific interaction events rather than running continuously) and epistemic in their function (i.e., targeting knowledge construction processes rather than behavioral compliance). The operators may be enacted at multiple levels: as system-generated prompts in a genAI-integrated learning environment, as pedagogical interventions implemented by an instructor, or as self-regulatory strategies internalized by learners through metacognitive training. AIRIS does not prescribe a single implementation mode; rather, it identifies the functional regulatory requirements that any productive hybrid cognitive system must satisfy. A key feature of AIRIS is that regulation is triggered by event-driven mechanisms that correspond directly to the destabilizing mechanisms identified above. Three principal trigger classes can be distinguished from interaction traces.

First, calibration discrepancy triggers activate when confidence levels diverge strongly from performance outcomes, indicating misaligned monitoring processes. Operationalizing calibration discrepancy in genAI-mediated contexts requires disentangling genAI-assisted performance from independent performance. This can be achieved through the systematic inclusion of brief probe tasks (i.e., short, targeted questions requiring independent knowledge construction without genAI access) that provide a reference point against which confidence judgments can be calibrated. Second, delegation dominance triggers activate when users consistently consult genAI before articulating their own ideas, signaling erosion of internal generative initiation. Third, reflection compression triggers activate when users move rapidly from genAI output to task completion without evaluation, indicating truncated regulatory cycles.[3] In SRL terms, these triggers collectively operationalize the monitoring function. They detect discrepancies between actual and desired epistemic engagement and initiate regulatory control. Thus the AIRIS architecture externalizes monitoring to behavioral traces, making regulatory triggers observable even when learners themselves are unaware of emerging instabilities. This externalization of monitoring raises a theoretically important caveat: AIRIS-triggered regulation should be understood as a scaffold, not a substitute, for the learner's own regulatory awareness. The long-term goal is not permanent external monitoring, but the internalization of regulatory competencies. Learners who recognize destabilization signals themselves and apply AIRIS operators as self-regulation strategies. AIRIS employs five complementary regulatory operators: *Anticipate*, *Interrogate*, *Reflect*, *Integrate*, and *Synthesize*. Those operators are flexible interventions that respond to specific diagnostic triggers rather than to a rigid sequence. The naming and structure of the five operators intentionally resonate with inquiry-based learning frameworks, in which learners generate hypotheses, pose questions, evaluate evidence, and construct meaning through iterative cycles (Hmelo-Silver et al., 2007; Pedaste et al., 2015). This resonance is deliberate: inquiry-based learning research has established that these functional steps are essential for deep knowledge construction, independent of instructional format. However, inquiry-based learning is a source of nomenclature and structural resonance, not a theoretical foundation, and AIRIS does not reduce to an inquiry-based learning model. The operators are grounded in three complementary theoretical traditions simultaneously: in resource terms, they keep constructive processing and the human–genAI division of labour active rather than allowing substitution (i.e., Anticipate, Interrogate); in representational terms, they sustain the active coordination of multiple representational formats (i.e., Integrate, Synthesize); and in regulation terms, they stabilize the forethought, monitoring, and reflection phases that genAI tends to compress (i.e., Anticipate, Reflect). Thus, inquiry-based learning provides one important lineage for the operator logic, but the operators apply across domains and task types far beyond the natural science inquiry contexts with which that tradition is most closely associated, including argumentative writing, legal reasoning, historical analysis, and professional judgment.

[3] The term reflection compression specifically refers to the temporal truncation of evaluative engagement - the absence of a genuine pause between receiving GenAI output and acting on it - rather than merely to the brevity of any particular reflection episode.

The Anticipate operator reinstates internal generative engagement. When delegation patterns indicate premature reliance on genAI, Anticipate prompts independent articulation of hypotheses, predictions, or goals. From a CLT perspective, this intervention ensures constructive processing remains active rather than being displaced by automated reasoning. Within SRL, it stabilizes the forethought phase that guides subsequent strategy selection. This operator draws on the tradition of inquiry-based learning (Hmelo-Silver et al., 2007), in which hypothesis generation prior to information gathering is considered as essential to active knowledge construction. In genAI contexts, Anticipate reinstates this generative step, which unrestricted access to genAI tends to bypass, as a structural prerequisite for productive consultation.

The Interrogate operator involves iterative interaction with genAI to acquire, refine, and compare candidate representations. The emphasis is not on asking questions per se, but on strategically using genAI as a representational partner while maintaining epistemic responsibility for the resulting outputs. Rather than outsourcing entire reasoning processes through unrestricted delegation, the Interrogate operator treats genAI consultation as a structured, recursive exchange in which the learner probes, contrasts, and revises candidate explanations, models, or solutions, and remains accountable for evaluating which of them to adopt. This aligns with hybrid intelligence perspectives that emphasize the complementarity between human and artificial capabilities (Dellermann et al., 2019; Salomon et al., 1991).

The Reflect operator re-establishes monitoring and evaluative engagement through justification prompts, requests for counterarguments, or explicit confidence judgments followed by verification. Reflection is essential for mitigating fluency effects that accompany polished genAI explanations. By requiring users to interrogate reasoning steps rather than simply accepting them, the Reflect operator strengthens calibration and counteracts monitoring compression.

The Integrate operator requires learners to connect validated genAI-generated outputs to prior knowledge, alternative representations, and domain conventions. Its function is to establish coherence across representations and resolve inconsistencies before further use. Whereas Reflect establishes whether a genAI contribution withstands scrutiny, the Integrate operator is the epistemic coordination of those validated representations: the learner relates the contribution to existing knowledge, reconciles it with competing representations, and resolves inconsistencies before the output is put to any further use. This operator draws on the multimedia-learning principle that meaningful learning depends on connecting new information to existing knowledge in long-term memory rather than appending it as an isolated fragment (Mayer, 2021), and on the meta-representational requirement that learners reconcile externally supplied representations with domain conventions (Ainsworth, 2006; diSessa, 2004). Therefore, the Integrate operator is where validation against domain conventions becomes binding: under conditions of recurring, domain-relevant genAI fallibility, a contribution that has been interrogated must be actively coordinated with the learner's representational repertoire, or rejected, rather than passively retained.

The Synthesize operator requires learners to transform validated and integrated outputs into a new epistemic product; its purpose is not integration but productive reconstruction. Whereas the Integrate operator coordinates representations to establish coherence, the Synthesize operator puts the coordinated representations to productive use, for example, generating explanations, constructing arguments, creating diagrams, or applying insights to novel contexts. In this sense, the Synthesize operator also counteracts representational laziness. genAI rapidly produces explanations, arguments, diagrams, summaries, figures, videos, and code fragments, yet their availability does not guarantee productive reconstruction. The Synthesize operator requires users to actively transform validated and integrated genAI outputs through generative processing rather than accepting them as-is. Depending on the domain, this may involve explaining a generated diagram verbally, paraphrasing a generated argument in one's own words, converting textual explanations into visual representations, identifying gaps or counterarguments in a generated essay, or applying a generated procedure to a novel case. The key is not the specific mode of transformation but the requirement for generative, productive reconstruction. This operator derives directly from multiple external representations and multimedia learning research,

To make the operators concrete, consider a single task traced through all five: a learner working to understand why a projectile's range is greatest at a launch angle of 45°. Anticipate has the learner first commit to a prediction and a rationale, perhaps reasoning that 45° balances horizontal and vertical velocity components, before any genAI consultation. Interrogate then uses genAI to probe that reasoning: the learner asks it to derive the range equation and compares the derivation against their own account, rather than requesting the answer outright. Reflect prompts the learner to check the returned derivation against their prediction and to notice discrepancies, for instance, that the range depends on $\sin(2\theta)$, which their intuitive account had not made explicit. Integrate requires coordinating the algebraic result with a graphical representation, sketching range as a function of angle and locating the maximum. Synthesize closes the cycle by having the learner produce a new artifact - a short written explanation, in their own words, of why the maximum falls at 45° - rather than accepting the genAI's explanation as finished. The example illustrates the general point: each operator reinstates a generative step that unrestricted genAI use would otherwise absorb.

The seven mechanisms and the five operators stand in a deliberate diagnostic–regulatory relationship: the mechanisms constitute the diagnostic side of the framework, specifying where and how hybrid cognition loses its equilibrium, whereas the operators constitute AIRIS's restabilizing layer, specifying the regulatory moves that restore it. Therefore, the operators are not a loose toolkit but the structural complement of the mechanisms - each operator names a generative or evaluative function whose attenuation a mechanism describes, so that to specify a mechanism is already to imply the operator that counters it. Table 1 summarizes how the five operators map onto the seven destabilizing mechanisms, indicating for each mechanism the operator(s) that most directly counteract it. The mapping is functional rather than exclusive: operators frequently co-occur, and a single mechanism may be addressed by more than one operator.

*Table 1. Functional mapping of the seven destabilizing mechanisms onto the five AIRIS operators.*

| **Destabilizing mechanism** | **Primary operator(s)** | **Regulatory rationale** |
|---|---|---|
| 1. Control redistribution | Whole AIRIS architecture (esp. Anticipate) | As the foundational mechanism, control redistribution is addressed by the architecture as a whole; Anticipate is the most direct first response, reinstating internal generative initiation before control shifts to genAI. |
| 2. Germane effort attenuation | Anticipate, Synthesize | Restores constructive, schema-building processing rather than passive reception. |
| 3. Calibration drift | Reflect | Re-establishes monitoring and confidence–accuracy alignment through evaluation. |
| 4. Delegation drift | Anticipate, Interrogate | Counters premature, default delegation by structuring when and how genAI is consulted. |
| 5. Representational passivity | Integrate, Synthesize | Forces active coordination of representations and their transformation into new products. |
| 6. Dual-threshold instability | Anticipate, Reflect | Maintains both internal engagement and monitoring so neither falls below its critical threshold. |
| 7. Motivational-affective drift | Anticipate, Synthesize | Preserves mastery and curiosity-based motivation by retaining effortful, self-authored work. |

*Note.* Assignments indicate the operator(s) most directly implicated; operators are not mutually exclusive and typically operate in combination.

### 3.3 Behavioral and Process Traces-Level Operationalization

Process-level learning analytics (i.e., methods that capture not just outcomes but the temporal dynamics of cognitive activity) provide the methodological infrastructure to study these regulatory dynamics empirically. Sequences of interaction data (i.e., task transitions, revision patterns, discourse structures, prompt sequences, and response latencies) can reveal how generative effort shifts between internal cognition and external systems (Gašević et al., 2017; Matcha et al., 2020). These developments create the possibility of linking theoretical insights about regulation and distributed cognition with empirical modeling of hybrid reasoning processes. For the AIRIS architecture to guide empirical research and system design, its mechanisms must be operationalized through observable behavioral indicators. The following operationalizations describe both empirical research strategies for testing AIRIS mechanisms and functional requirements for technical implementations of the Detect-Diagnose-Act cycle. They are theoretical proposals; concrete instantiations will necessarily depend on the specific task, domain, and assessment context. Recent developments in learning analytics, process mining, and human-AI interaction modeling provide precisely the methodological tools required (Gašević et al., 2017; Matcha et al., 2020; Winne, 2017).

Control redistribution can be indexed through behavioral proxies, such as the latency to first genAI consultation, the proportion of independently generated content, and longitudinal reduction in independent planning, with sequence modeling reconstructing whether episodes begin with independent generation or with genAI followed by minimal human modification (Matcha et al., 2020).

Germane effort attenuation can be approximated through semantic transformation analysis (e.g., semantic distance or paraphrase originality via NLP), where persistently low transformation across complex tasks signals recognition-based rather than generative revision.

Calibration dynamics can be examined through confidence-accuracy modeling: structured confidence judgments before and after genAI exposure yield calibration slopes, and progressively weaker slopes under unrestricted use would indicate monitoring compression, with verification-action frequency providing a complementary trace indicator.

The dual-threshold instability prediction is the most theoretically novel and empirically demanding of the AIRIS propositions: as a claim about nonlinear interaction effects, it requires factorial designs that jointly manipulate internal engagement and genAI reliability across well- and ill-structured problems, predicting disproportionate degradation under joint insufficiency rather than additive error accumulation. These analytic and experimental approaches collectively point toward the possibility of closed-loop hybrid regulation systems in which AIRIS functions as both a theoretical framework and a computational architecture. Interaction traces would first be analyzed to detect instability markers; diagnostic models would classify the underlying mechanism; the system would then activate the appropriate operator. Such Detect-Diagnose-Act architectures align with calls to move learning analytics beyond descriptive dashboards toward actionable regulation systems (Gašević et al., 2017).

To indicate how these propositions translate into observable research targets, Table 2 lists, for several core constructs, example measures that prior work has used or that follow directly from the construct definitions. The list is illustrative rather than exhaustive; its purpose is to show that the framework's constructs are empirically tractable and that the AI-withdrawal logic, assessing competence once genAI support is removed, illustrates a unifying measurement strategy whose domain-specific operationalization and psychometric validation lie beyond this framework.

*Table 2. Illustrative measurement targets for selected AIRIS constructs.*

| Construct | Example measures |
|---|---|
| Calibration compression (Mechanism 3) | Confidence–accuracy alignment (calibration curves, bias/discrimination indices) |
| Delegation drift (Mechanism 4) | genAI-usage patterns (frequency/latency of queries, proportion of task delegated, log-based traces) |
| Representational passivity (Mechanism 5) | Transfer performance after genAI withdrawal (independent problem solving, representation translation tasks) |
| Germane effort attenuation (Mechanism 2) | Invested mental effort and depth-of-processing indicators (self-report effort, think-aloud, elaboration coding) |
| Motivational-affective drift (Mechanism 7) | Intrinsic motivation, interest, and effort valuation (questionnaires; behavioral persistence on effortful tasks) |

*Note.* Transfer after genAI withdrawal serves as a cross-cutting indicator of whether epistemic agency has been preserved.

## 4. Epistemic Agency as Normative Benchmark

The central normative construct of AIRIS is *epistemic agency*. As introduced above, this benchmark is normative but empirically anchored. Epistemic agency, as defined here, is domain-general, and the present framework is deliberately confined to that general, domain-independent level. How epistemic agency specializes to particular domains, and how such domain-specific forms are defined and measured, lies beyond the scope of this framework. This general orientation has received independent theoretical support. Brod (2026) argued that agency in AI-mediated learning environments cannot be equated with choice: learners may freely decide how to use AI but their choices are structurally pre-formed by efficiency incentives, interface design, and habituated delegation patterns. This argument applies with particular force to genAI contexts: when high-quality outputs are consistently available at low effort, the decision architecture systematically rewards delegation over engagement. The deeper concern is not merely that genAI makes disengagement attractive, but that sustained disengagement may undermine the very preconditions for good epistemic decision-making, much as habitual, frictionless consumption can erode the capacities it depends on. If learners rarely exercise generative initiation, tolerance for productive struggle, and self-monitoring, the capacity to judge when to rely on genAI at all may itself atrophy, making the “choice” to delegate progressively less free. Over time, this may not only reshape behavioral patterns but reduce the cognitive capacities, the habit of generative initiation, the tolerance for productive struggle, the skill of self-monitoring,that epistemic agency requires. Freedom of action does not constitute epistemic agency when the decision architecture systematically rewards disengagement. AIRIS operationalizes this insight by targeting not the availability of genAI but the regulatory conditions under which genAI use either preserves or erodes internal generative participation.

As defined in Section 1, epistemic agency is the capacity to initiate, monitor, validate, coordinate, and take responsibility for one’s own knowledge-generating processes. This concept is distinct from behavioral autonomy: a user may freely choose to rely heavily on genAI while engaging minimally in generative reasoning, exercising freedom without epistemic engagement. It is also distinct from performance outcomes: high-quality outputs may emerge even when individuals contribute little to the underlying generative reasoning. In hybrid systems, the critical issue is not simply what knowledge products are produced but how cognitive control over generative processes is exercised.

AIRIS introduces a theoretically important distinction between output ownership and generative ownership. Individuals may be credited with producing a solution, whereas the underlying reasoning is largely generated by an external system. Epistemic agency depends on generative ownership, i.e., the active construction and evaluation of reasoning processes, even when external tools contribute to the generation of outputs. This distinction has direct implications for how genAI-assisted work should be evaluated in educational, professional, and scientific contexts.

Metacognitive calibration represents a critical dimension of epistemic agency. Accurate self-assessment is essential not only for learning but for responsible professional judgment. GenAI systems complicate calibration by providing fluent outputs that may inflate perceived comprehension. Therefore, the AIRIS framework treats calibration maintenance as a primary regulatory goal rather than merely an outcome measure. A critical distinction for epistemic agency concerns the difference between performance goals (producing high-quality immediate outputs) and learning goals (developing durable competencies). genAI systematically privileges the former: it optimizes for output quality while potentially undermining the cognitive engagement required for the latter. Calibration maintenance within AIRIS specifically concerns the learning-goal dimension: do individuals accurately assess whether they have understood what they were working on, not merely whether they have produced a satisfactory output?

Beyond cognitive calibration, genAI may also affect motivational calibration, the accuracy with which individuals assess their own interest, curiosity, and engagement. When genAI systems consistently frame topics as fascinating and important, learners may lose the ability to distinguish genuine intrinsic interest from geAI-induced enthusiasm (Murayama, 2022). This motivational dimension of calibration has received less attention in the genAI literature but may be equally consequential for sustaining long-term epistemic engagement.

Contemporary human-centered AI frameworks emphasize transparency, interpretability, and human oversight (Shneiderman, 2022). Yet oversight alone does not guarantee epistemic agency. A user who merely reviews genAI outputs without engaging in generative reasoning exercises formal oversight while remaining cognitively disengaged. Epistemic agency requires participatory construction rather than passive supervision. Consider the analogy of a journal peer reviewer: both a deeply engaged reviewer (who reconstructs the argument, identifies gaps, proposes alternatives) and a superficial reviewer (who checks for plausibility without genuine engagement) exercise formal oversight. Only the former exercises epistemic agency. genAI users face an analogous choice, and an analogous temptation toward the latter. AIRIS operationalizes this principle by ensuring that generative participation, reflective evaluation, and representational integration remain active components of hybrid cognition.

## 5. Theoretical Propositions and Empirical Program

AIRIS is intended not merely as a descriptive synthesis but as a theoretical framework capable of producing falsifiable predictions about hybrid human-AI cognition. Accordingly, predictive fit alone is insufficient for mechanistic claims; these require construct alignment and causal intervention tests (Wang et al., 2026). The following six propositions articulate interlocking empirical claims derived from the mechanisms identified above.

(1) *Longitudinal delegation drift*: Repeated interaction with genAI systems gradually reshapes problem-solving strategies. Users who initially consult genAI only after independent attempts at reasoning will increasingly rely on it earlier in the problem-solving process. This predicts that independent generative initiation will decline across repeated interactions, even when immediate performance improves. Indicators include decreasing latency to first genAI query, reduced independent planning articulation, and lower diversity of independently generated strategies (Bastani et al., 2025). Direct experimental evidence is consistent with this prediction, though it does not yet track the drift longitudinally: in randomized withdrawal experiments, Liu et al. (2026) found that even brief genAI-assisted problem solving left learners performing worse once assistance was removed and abandoning tasks more often than controls, with these costs concentrated among those who had used genAI primarily to obtain direct solutions.

(2) *Generative engagement mediates transfer*: Internal generative engagement mediates the relationship between genAI-supported performance and long-term learning outcomes. When genAI assistance replaces rather than scaffolds constructive processing, immediate performance may improve without corresponding gains in transferable understanding. Conditions that enforce generative initiation and representational synthesis (via the Anticipate, Integrate, and Synthesize operators) should yield stronger transfer and retention outcomes than unrestricted reliance on genAI.

(3) *Calibration drift under unrestricted genAI use*: Calibration slopes (i.e., quantifying the confidence-accuracy relationship) will decline across repeated genAI interactions in unregulated environments but remain stable when structured reflection mechanisms (Reflect operator) are embedded in the interaction process. The fluency of genAI reasoning creates an illusion of understanding that progressively weakens monitoring coherence unless reflective evaluation is explicitly enforced.

(4) *Dual-threshold instability*: Hybrid performance depends jointly on sufficient internal generative engagement and sufficient genAI reliability relative to task complexity. Dual-threshold instability predicts disproportionate increases in error, specifically when both fall below task demands simultaneously. That is, a nonlinear interaction pattern emerges, reframing hybrid failure as a relational rather than additive phenomenon.

(5) *Representational synthesis as a stabilizing mechanism*: Conditions requiring learners to translate genAI outputs across representational formats will produce stronger retention and transfer outcomes than conditions in which representations are simply consumed. Representational synthesis preserves constructive depth within hybrid reasoning systems and maintains the active integration essential to durable understanding.

(6) *Motivational recalibration*: Sustained exposure to frictionless genAI-supported problem solving reduces voluntary engagement with demanding generative reasoning. Behavioral indicators (e.g., voluntary engagement with optional challenge tasks, persistence under difficulty) should show longitudinal decline under frictionless genAI conditions, consistent with motivational-affective drift, reinforcing delegation and disengagement cycles.

These propositions form a coherent multi-level causal chain: delegation drift reduces constructive engagement; weakened engagement produces calibration drift; calibration drift increases vulnerability under dual-threshold conditions; representational passivity limits conceptual integration; and motivational drift amplifies and sustains these dynamics. Each link in this chain can be tested independently through experimental designs, longitudinal modeling, and trace-based learning analytics, thereby constituting a cross-disciplinary research agenda for regulated human-AI collaboration.

## 6. General Discussion

### 6.1 Contributions

AIRIS advances theory in several ways that collectively reframe the study of genAI augmentation, recasting long-standing constructs, cognitive load, constructive processing, self-regulation, and representational competence, as components of a single regulatory question about where generative and evaluative control reside. First, AIRIS moves beyond trust and technology acceptance models toward a structural account of generative participation. Classical frameworks address whether individuals rely on genAI systems; AIRIS addresses how cognitive control is redistributed during generative reasoning, which is a qualitatively different question with distinct empirical implications.

Second, AIRIS integrates three previously separated theoretical traditions: (a) resource-based accounts (cognitive load theory, distributed cognition, and cognitive offloading), which explain how cognitive load is distributed between learner and genAI and how working-memory constraints shape learning; (b) representation-based theories (CTML, MERs/DeFT), which explain how learning depends on the active coordination of multiple representational formats; and (c) regulation-based theories (SRL), which explain how learners monitor and control their own cognitive processes. AIRIS demonstrates that all three are necessary: resource allocation, representational integration, and regulatory monitoring must be considered jointly to understand and address the full range of destabilization mechanisms. AIRIS bridges this gap by embedding external generative support within regulatory cycles and demonstrating that stability depends on maintaining coherence among monitoring, control allocation, and representational integration.

Third, AIRIS formalizes interactional instability in hybrid reasoning systems. Traditional frameworks treat human and algorithmic errors as additive. AIRIS proposes that hybrid failures are relational, emerging when both internal generative engagement and genAI reliability fall below task demands. This dual-threshold formulation aligns with hybrid intelligence theory (Dellermann et al., 2019) while offering a more precise account of when hybrid collaboration becomes unstable. Fourth, AIRIS positions epistemic agency, not efficiency, as the central evaluative benchmark for hybrid cognition. By introducing the distinction between output ownership and generative ownership, the framework clarifies why performance metrics alone are insufficient for evaluating genAI-assisted work. Fifth, through its Detect-Diagnose-Act architecture, AIRIS connects empirical analytics with theory-driven design, providing a pathway from observable interaction traces to targeted regulatory interventions and thereby moving learning analytics beyond description toward actionable support (Gašević et al., 2017; Shneiderman, 2022).

Although grounded partly in educational theory, AIRIS extends to other knowledge-intensive domains, scientific, legal, medical, and software work (Birhane et al., 2023;

Dell'Acqua et al., 2023; Peng et al., 2023), where the challenge is likewise maintaining human understanding and accountability within hybrid reasoning rather than genAI accuracy alone. Finally, the present account is bounded: AIRIS is formulated for an individual learner working with genAI, whereas much epistemic work is collaborative. Extending the framework to group-level regulation is the most substantial development it requires, which we take up as a research direction in Section 6.2.

### 6.2 Future Research

In this section, we discuss various important research areas identified within the AIRIS framework. A first research area concerns the longitudinal dynamics of control allocation. The current literature on genAI predominantly focuses on short-term performance. AIRIS predicts that considerable cognitive effects emerge over longer interaction periods as delegation drift accumulates. Longitudinal designs that examine how generative participation evolves across repeated sessions, analyzed using process mining, hidden Markov models of control-state transitions, or growth curve modeling of trace indicators, are therefore essential.

A second area concerns individual differences in regulatory sensitivity. AIRIS predicts that delegation drift, calibration drift, and motivational-affective drift vary substantially across individuals as a function of metacognitive skill, domain expertise, self-efficacy, interest, and genAI experience. Understanding these moderating factors is essential for developing adaptive regulatory systems that respond to individual regulatory profiles rather than applying uniform interventions.

A third area concerns the collaborative and organizational dimensions of hybrid cognition. Multi-level designs that examine regulatory dynamics at the individual, dyadic, team, and organizational levels are needed. Research should investigate how genAI-dominant workflows affect epistemic culture within teams and whether AIRIS regulatory structures can preserve socially shared generativity at scale. This is where the framework most clearly requires extension. Research on socially shared regulation shows that effective collaboration depends on collective monitoring, explanation, and evaluation, and that shared regulatory responsibility already carries a risk of diffusion, in which members tacitly assume that monitoring is handled by others (Järvelä & Hadwin, 2013). Introducing genAI into such groups may sharpen this risk: when genAI outputs become the group's de-facto reference artifact, collaborative reasoning can shift from co-construction toward the collective evaluation of externally generated material, reducing peer-level generativity. Whether the five operators can be re-specified for group-level enactment, so that Anticipate, Interrogate, and the rest are performed jointly rather than individually, is, in our view, the single most important direction for extending AIRIS beyond the individual learner.

### 6.3 Design Implications

These research areas have corresponding design implications. genAI interfaces should incorporate prior articulation requirements: structured prompts that require users to formulate their own hypotheses, goals, or partial solutions before accessing genAI support. Reflective checkpoints before submission can restore calibration processes and prevent the truncation of evaluative engagement, the tendency to move directly from genAI output to task completion without genuine reflection. Representational translation tasks embedded within genAI-assisted workflows can preserve constructive depth by requiring active cross-modal integration. Uncertainty visualization, displaying genAI confidence estimates and alternative reasoning pathways, can support calibration by making epistemic limitations visible. The uncertainty indicator could also refer to the availability of training data, as well as its quality and quantity. Adaptive architectures capable of detecting instability markers and triggering targeted regulatory interventions can maintain stability dynamically.

A key design question for implementation concerns the degree of automation: whether AIRIS functions as a fully automated intervention system (detecting instability and triggering regulatory responses without user awareness) or as a transparent metacognitive coach (making destabilization signals visible to learners and supporting their own regulatory agency). The latter aligns more directly with AIRIS's normative commitment to epistemic agency: if the system regulates on the learner's behalf without fostering regulatory awareness, it risks substituting a new form of dependency for the one it is meant to prevent.

These design implications must be considered in light of ethical constraints. Trace-based monitoring systems should operate transparently, with users understanding when and why regulatory interventions occur. Interventions must scaffold rather than constrain autonomy. Data collection should remain minimal and clearly justified. AIRIS aligns with human-centered genAI principles emphasizing empowerment, transparency, and accountability (Shneiderman, 2022; Floridi et al., 2018).

Two further implications follow from evidence on how reliance develops. First, systems should make their own fallibility visible rather than presenting outputs with uniform confidence. K. Bauer et al. (2026) found that users who encountered a system's errors early calibrated their reliance appropriately and retained independent skill, whereas those who did not over-relied; deliberately surfacing uncertainty and occasional visible limitations may therefore protect calibration better than a seamless interface. Second, delegation should be designed to be reversible: because the drift documented here accumulates gradually and often unnoticed, interfaces that let users temporarily disable assistance, or that periodically require unaided performance, can counteract the ratchet by which each act of offloading makes the next more likely. Pi et al. (2026) point in the same direction at the level of task design, showing that having learners locate their own errors before genAI feedback yields larger gains than genAI-led correction. These implications extend beyond education: in professional settings such as clinical decision support, financial analysis, and software development, the same reliance dynamics apply, and the same design commitments, visible fallibility, reversible delegation, and prior human engagement, are correspondingly relevant.

Finally, trace-based regulation raises ethical considerations that bear directly on design: systems must remain transparent, interventions must scaffold rather than constrain autonomy, and data collection must adhere to privacy principles. Throughout, AIRIS prioritizes preserving epistemic agency over optimizing performance metrics alone.

## 7. Conclusion

Human-AI collaboration, especially human-genAI collaboration, has moved beyond tool use into epistemic co-production: genAI systems now generate hypotheses, draft arguments, propose explanations, and synthesize evidence alongside (and sometimes instead of) their human partners. Empirical studies document measurable productivity gains, yet performance alone is an insufficient metric for evaluating hybrid human-AI systems. The performance-metacognition dilemma (i.e., improved outputs alongside weakened generative depth) demands theoretical clarification that existing frameworks have not provided.

In this paper, we have argued that genAI integration fundamentally reconfigures cognitive control. genAI redistributes generative processes across internal and external loci, potentially enhancing efficiency while attenuating constructive processing, calibration accuracy, and motivational effort. From this analysis, we hypothesized seven interacting destabilizing mechanisms and derived AI-Augmented Inquiry and Regulation in Hybrid Systems (AIRIS) as a multi-level framework designed to stabilize hybrid generative cognition.

The central challenge of genAI-supported learning is no longer access to information or representations, but preserving epistemic agency when generative work becomes externally available. AIRIS reframes genAI augmentation as a dynamic equilibrium problem. Stability depends on maintaining sufficient internal generative engagement relative to task demands, monitoring-control coherence, representational integration, complementarity between bounded human and artificial agents, and motivational commitment to productive difficulty. The five AIRIS operators Anticipate, Interrogate, Reflect, Integrate, and Synthesize provide event-driven regulatory mechanisms that restore equilibrium once destabilizing patterns emerge. The account developed here is deliberately bounded: it describes an individual epistemic agent working with a generative system. Extending it to collaborative and institutional settings is the next substantial step.

Importantly, with AIRIS, we do not position genAI as a threat to cognition, nor do we advocate restriction of generative technologies. Rather, with AIRIS, we articulate conditions under which augmentation preserves rather than erodes epistemic robustness. Hybrid cognition is not inherently destabilizing; instability arises when control is redistributed without regulatory scaffolding. As genAI becomes embedded in scientific discovery, medical diagnosis, legal drafting, policy design, and educational assessment, preserving epistemic agency becomes a societal imperative. Systems optimized exclusively for output efficiency risk cultivating shallow generative participation and fragile reasoning structures. AIRIS offers a theoretically grounded architecture for identifying instability, diagnosing control imbalances, and restoring productive, generative engagement, providing a conceptual foundation for the next phase of human-AI collaboration.